# Transport and Phototransport in ITO Nanocrystals with Short to Long-Wave Infrared Absorption

Junling Qu[1], Clément Livache[1,2], Bertille Martinez[1,2], Charlie Gréboval[1], Audrey Chu[1], Elisa Meriggio[1], Julien Ramade[1], Hervé Cruguel[1], Xiang Zhen Xu[2], Anna Proust[3], Florence Volatron[3], Grégory Cabailh[1], Nicolas Goubet[1,2,4], Emmanuel Lhuillier[1*]

[1] Sorbonne Université, CNRS, Institut des NanoSciences de Paris, INSP, F-75005 Paris, France
[2] Laboratoire de Physique et d'Etude des Matériaux, ESPCI-Paris, PSL Research University, Sorbonne Université UPMC Univ Paris 06, CNRS, 10 rue Vauquelin 75005 Paris, France.
[3] Sorbonne Université, CNRS, Institut Parisien de Chimie Moléculaire, IPCM, F-75005 Paris, France.
[4] Sorbonne Université, CNRS, Laboratoire de la Molécule aux Nano-objets ; Réactivité, Interactions et Spectroscopies, MONARIS, F-75005 Paris, France

**Abstract**: Nanocrystals are often described as an interesting strategy for the design of low-cost optoelectronic devices especially in the infrared range. However the driving materials reaching infrared absorption are generally heavy metal-containing (Pb and Hg) with a high toxicity. An alternative strategy to achieve infrared transition is the use of doped semiconductors presenting intraband or plasmonic transition in the short, mid and long-wave infrared. This strategy may offer more flexibility regarding the range of possible candidate materials. In particular, significant progresses have been achieved for the synthesis of doped oxides and for the control of their doping magnitude. Among them, tin doped indium oxide (ITO) is the one providing the broadest spectral tunability. Here we test the potential of such ITO nanoparticles for photoconduction in the infrared. We demonstrate that $In_2O_3$ nanoparticles presents an intraband absorption in the mid infrared range which is transformed into a plasmonic feature as doping is introduced. We have determined the cross section associated with the plasmonic transition to be in the $1-3\times10^{-13}$ cm$^2$ range. We have observed that the nanocrystals can be made conductive and photoconductive due to a ligand exchange using a short carboxylic acid, leading to a dark conduction with n-type character. We bring further evidence that the observed photoresponse in the infrared is the result of a bolometric effect.

To whom correspondence should be sent: el@insp.upmc.fr

## INTRODUCTION

Over the recent years, huge progresses[1,2,3,4] have been made in the field of infrared (IR) optoelectronics using colloidal quantum dots as a low-cost alternative to historical semiconductor technologies (InGaAs, InSb, HgCdTe…). In the visible range, InP[5,6] has appeared as viable alternative strategy to Cd-based nanocrystals for light emission. However, the infrared range is still driven by toxic compounds and more particularly lead[7,8] and mercury chalcogenides[9,10].

Since materials above arise a toxicological concern for mass-market applications, it is of utmost importance to screen alternative materials with a reduced toxicity and an optical absorption in the mid infrared. However, in practice narrow band gap is often found with heavy atoms which are more likely to be toxic. In this sense, doped semiconductors presenting intraband transition in the range of wavelength of interest need to be considered. Mercury chalcogenides Self-doped nanocrystals[11] present a tunable intraband absorption[12,13,14] in the 3 to 60 µm range[15]. $Ag_2Se$[16,17,18], which is heavy metal-free, presents a very similar absorption spectrum to HgSe, but its performance for photodetection remains far weaker than its Hg-containing counterpart[19].

Degenerately doped oxides[20,21,22] such as Ga[23], Al[24] and In[25] doped ZnO, indium tin oxide (ITO)[26,27,28,29], tungsten oxide[30] and cadmium oxide have seen prompt advances in terms of synthetic maturity over the recent years. Compared with the doping of II-VI and IV-VI materials which remains difficult and mostly induced through surface effects, the doping of oxides seems more straightforward and tunable. Like bulk semiconductors, the introduction of heteroatoms leads to the generation of free carriers, resulting in plasmonic mid infrared absorption and carrier conduction.[31,32] The optical absorption of such doped oxides has been extensively discussed, in particular to understand the correlation between the plasmonic absorption linewidth[33] and dopant position.[34,35,36,37] On the application side, such doped oxides have been driven by their use in the field of smart windows,[38] sensing[39] and catalysis.[40] However, their photoconductive performance remains nearly unexplored. In a semiconductor, the photocurrent is proportional to two key material parameters which are material absorption and photocarrier lifetime. For interband and intraband transitions, the absorption cross section per nanoparticle is quite weak, typically in the $10^{-15}$ $cm^2$ range.[41,42] This implies that for an intraband transition from a lightly doped semiconductor (1 carrier per dot), only one electron is active every $10^3$-$10^4$ atoms. On the other extreme, metal with 1 active carrier per atom can reach high cross section[43] ($10^{-12}$ $cm^2$ per particle). The two type of transitions also differ by the fact that intraband absorption is quantum confined (*ie* size depedent) while plasmonic absorption is a bulk like property. In a degenerately doped semiconductor, the situation is typically intermediate, but the cross section is higher than that of marginally doped materials. The drawback of plasmonic absorption comes from its shorter lifetime compared with exciton which may limit the benefit of a higher absorption. Thus, the exact potential of oxide nanocrystals for IRphotoconduction remains to be determined. In addition, several open questions have to be tackled such as the exact nature of the transition in the low doping regime and the IR photoconduction potential. In particular, the electrical nature (photoconduction vs bolometric) of the response under IR illumination needs to be revealed. In this paper, we probe the transport properties of ITO nanoparticles both in dark condition and under illumination to answer the previously mentioned questions. It is also of utmost importance to quantize the IR photoresponse performance of these ITO nanocrystals with the one obtained using undoped or barely doped excitonic semiconductors.

## DISCUSSION

We start by synthetizing a series of ITO nanocrystals with various levels of doping from 0 ($In_2O_3$) to 15% of Sn content using a previously reported method.[31] The nanocrystal solution, see Figure 1 a, shows a color switching from colorless for undoped material to blue for doped material. The nanoparticles are of similar size (≈12 nm) with a near sphere shape for all doping levels, as revealed by transmission electron microscopy (TEM), see Figure 1 b and c and S1. The X-ray diffraction, see figure S2, confirms the bixbyite structure of the $In_2O_3$ lattice. These nanoparticles present an absorption peak in the infrared as well as a band edge in the UV range, see Figure 1d. Both transitions blueshift with the increase of Sn content. The band edge transition shift is consistent with the bleach of the lowest energy state as the conduction band state get filled, see figure S3. The infrared peak shifts from 9.3 µm (≈148 meV) for the undoped material to 2 μm (≈614 meV) for the 10% doped particles, see Table 1. We observe that above 10% the transition barely blueshifts, but strongly broadens due to an increase of electron-electron scattering.[37] In the following, we will focus more specifically on undoped ($In_2O_3$), 1.7% Sn and 10% Sn doped ITO nanoparticles.

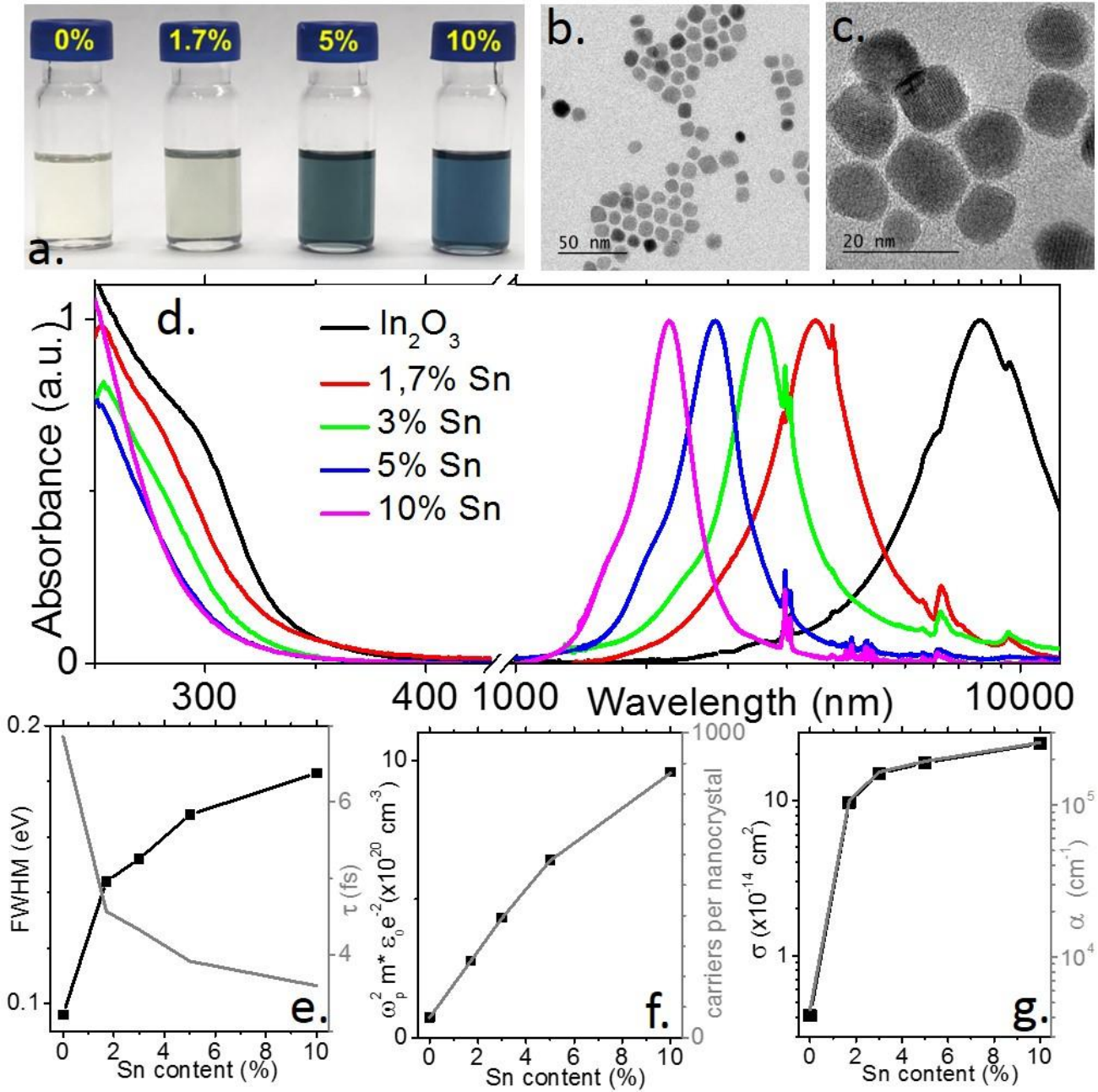

Figure 1a. Image of ITO nanoparticles dispersed in $CHCl_3$ with various levels of Sn content. b. Transmission electron microscopy image of the 1.7% Sn ITO nanocrystals. c. High-resolution TEM image of the 1.7% Sn ITO nanocrystals. d. Infrared and UV-visible spectra of ITO nanocrystals with various levels of doping. e. Full width at half maximum (FWHM) of the infrared absorption and the estimated average electron scattering time τ (ℏ/FWHM) as a function of the Sn content. f. Absorption cross section per nanoparticle and film absorption coefficient as a function of the Sn content. g. Estimated carrier density from the plasmon peak energy per unit of volume and per nanoparticle as a function of Sn content.

The infrared peak has been assigned to plasmonic absorption, which allows us to determine the carrier density. From the Drude model[44], one can relate the localized surface plasmonic resonance frequency ($\omega_{LSP}$) to the bulk plasma frequency ($\omega_p$) via $\omega_{peak} = \sqrt{\frac{\omega_p^2}{\varepsilon_\infty + 2\varepsilon_M} - \gamma^2}$ where $\varepsilon_\infty = 4$ is the high frequency dielectric constant of the ITO nanoparticles[36], $\varepsilon_M \approx 2$ the dielectric constant of the environment and $\gamma$ the plasmon damping constant. The latter is estimated as the linewidth of the plasmon transition, *i.e.* the full width at half maximum (FWHM) of the transition. The average scattering time $\tau$ can be estimated from $\gamma = \hbar/\tau$, with $\hbar$ the reduced Planck constant. The value of $\tau$ has been estimated to be ≈ 5±1 fs, see Figure 1e. The plasma frequency directly relates to the carrier density (*n*) through the relation $\omega_p^2 = \frac{ne^2}{\varepsilon_0 m^*}$ with *e* the proton charge, $\varepsilon_0$ the vacuum dielectric constant and *m\** the ITO conduction band effective mass, here taken equal to $0.4 m_0$ [45,34] with $m_0$ the electron rest mass.

*Table 1 Band edge energy, energy of the infrared absorption peak and absorption cross section associated with the infrared transition for three Sn content of ITO nanoparticles.*

| Sn content | Band edge energy (eV) | Infrared peak energy (meV) | Infrared cross section (x$10^{-15}$ cm²) |
|---|---|---|---|
| 0 % | 3.8 | 148 (8.3 µm) | 4 |
| 1.7 % | 4 | 315 (3.95 µm) | 90 |
| 10 % | 4.3 | 614 (2.02 µm) | 220 |

We estimate the carrier density to be in the $10^{20}$-$10^{21}$ cm$^{-3}$ range, which corresponds to 250 (resp. 900) free electrons in the case of the 1.7% doped ITO nanoparticles (resp. 10%), see Figure 1f. In the case of the undoped material, this corresponds typically to 100 electrons per nanocrystal, however the validity of the Drude model is questionable in this case, as discussed later in the text.

We also have determined the absorption cross-section of the nanoparticles as a function of doping, see Figure 1g, Table 1 and Methods. Doped nanoparticles have a cross section in the 1-3x$10^{-13}$ cm² range. While the cross section increases quasi linearly with doping for doped nanoparticles, the undoped nanoparticles present a much weaker absorption with a cross section of only 4x$10^{-15}$ cm². This is a first indication that the number of free carriers in undoped nanoparticles can be much lower than the one predicted by the Drude model.

Transport measurements are used to probe the carrier density. To ensure a strong interparticle coupling in the film, a ligand exchange step is required. While thiols have been extensively used for Cd, Hg or Pb based nanocrystals, the hard base nature of In$^{3+}$ leads to a poor affinity for thiol. As a result, we choose a hard acid, according to Pearson's theory, such as acetic acid to replace the initial oleic acid ligands. The transport properties of the ITO nanocrystals under dark condition are probed using an electrolyte field effect transistor configuration, see a scheme of the setup in Figure 2a. Electrolyte gating, here made by dissolving LiClO$_4$ in a polyethylene glycol matrix, ensures a large gate capacitance which is critical to modulate the carrier density of degenerately doped material. In addition, it allows air operation and gating of thick films[46].

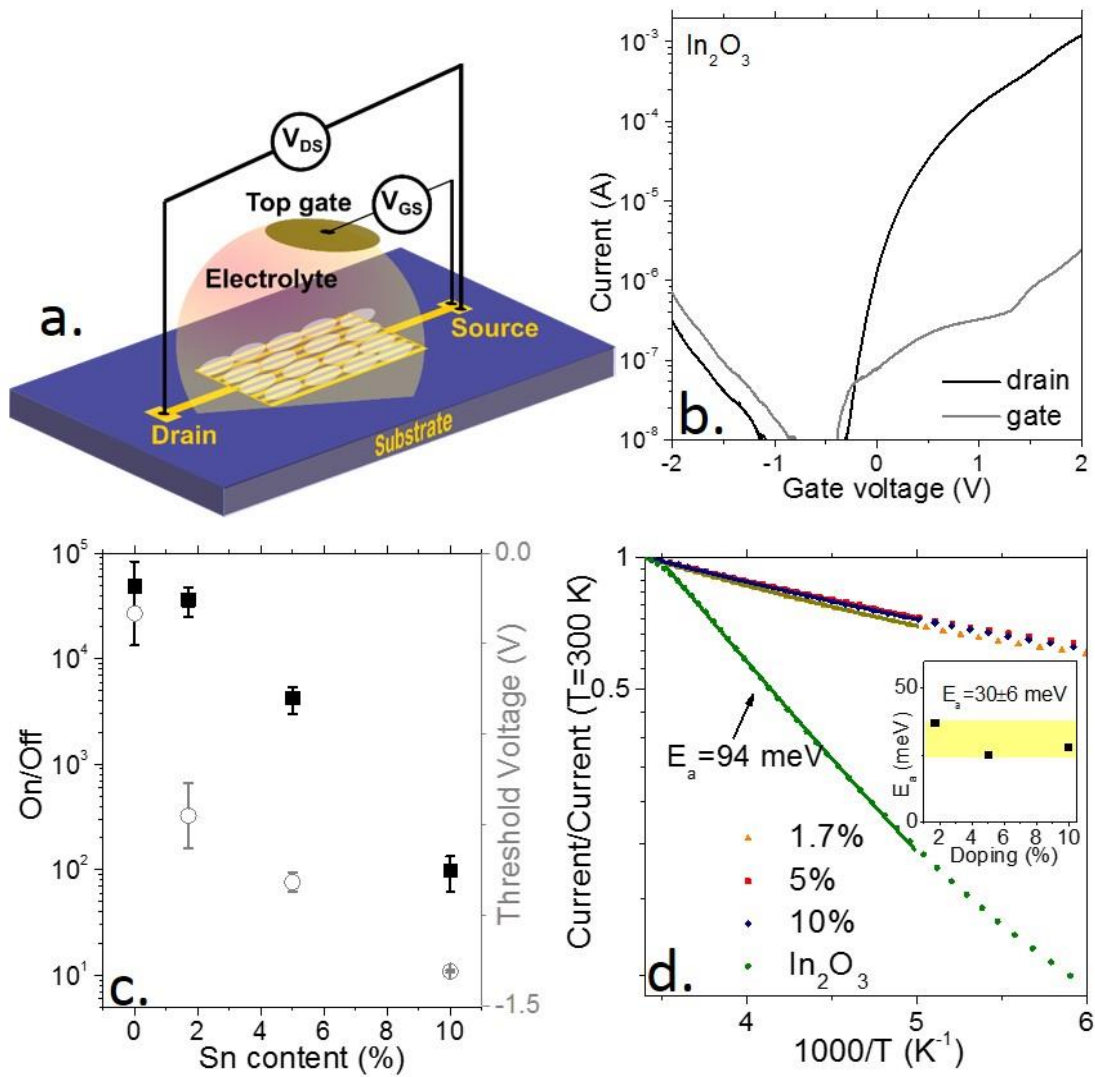

*Figure 2a. Scheme of an electrolyte gated transistor with a thin film of ITO nanocrystals as canal. b. Transfer curve (drain and gate current as a function of the applied gate voltage) for a transistor whose canal is made of $In_2O_3$ nanocrystals. c. On/off ratio and threshold voltage as a function of the Sn content for a transistor made of ITO nanocrystals thin film. d. Current as a function of temperature for ITO nanocrystal thin films with various Sn contents. The inset provides the fitted activation energy for the doped samples assuming an Arrhenius fit of the high temperature part of the I-T curve.*

Thin films of ITO and undoped $In_2O_3$ nanoparticle only present n-type conduction (*i.e.* rise of conduction under positive gate bias), see Figure 2b and Figure S4. No evidence for hole transport has been observed, which is consistent with the wide band gap nature of ITO.[46] We nevertheless observe a clear reduction of the current modulation (on/off ratio) with the Sn content increase as well as a shift of the threshold voltage toward more negative potentials, see Figure 2c. Both are consistent with the increase of doping leading to a Fermi level deeper within the conduction band. Around room temperature, the transport is thermally activated, see Figure 2d. An activation energy of almost 100 meV in the case of $In_2O_3$ is extracted from the Arrhenius fit of the I-T curve. This value is weaker for doped samples at 30±6 meV, independent of the doping level, see the inset of Figure 2d. The latter value can be related to the charging energy ($E_C$) of the nanoparticle where $E_C$ is given by $E_C = \frac{e^2}{2C}$ with C the self-capacitance of the nanoparticle. C can be estimated from $C = 2\pi\varepsilon_0\varepsilon_r d$ with $\varepsilon_0$ the vacuum permittivity, $\varepsilon_r$ the material dielectric constant (4) and $d$ the nanoparticle diameter (12 nm), which leads to 29 meV. In this sense, the driving energy of the hopping transport in the array of degenerately doped semiconductors is the same one as for metallic nanoparticles.[47]

While dark conduction in ITO nanoparticle film has already been explored,[31,32,48] their photoconductive properties remain mostly unexplored. We choose to study their photoconductive properties by exciting them at the band edge

(UV irradiation) or by directly exciting their plasmonic feature in the near-IR for 10% doped and in the mid-IR for the 1.7% doped nanoparticles. In practice, we use four light sources to resonantly excite each of these transitions, see Figure 3a.

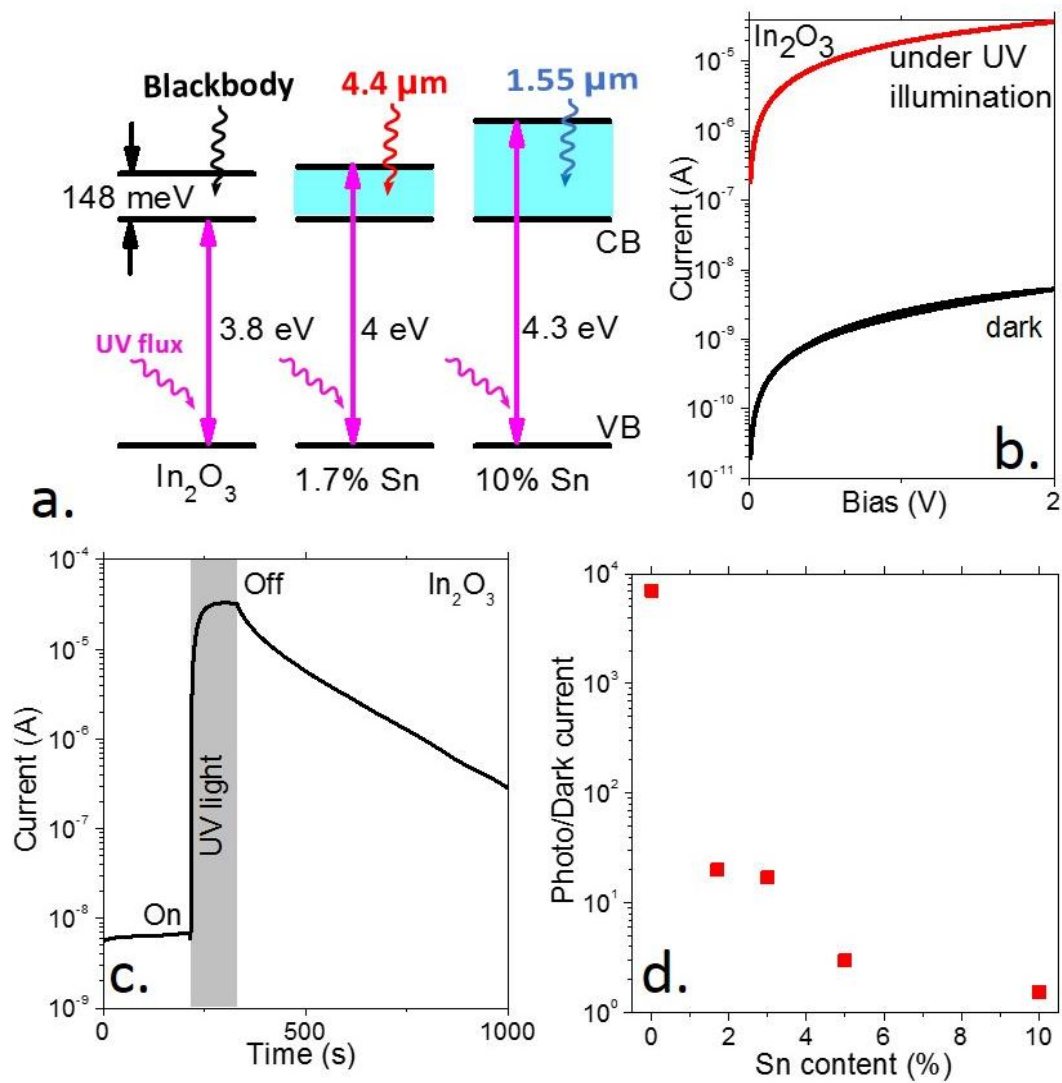

Figure 3a. Band diagram of undoped, 1.7% Sn and 10 % Sn doped ITO nanocrystals. b. I-V curve of $In_2O_3$ nanoparticles under dark condition and under UV illumination. c. Current as a function of time while a thin film $In_2O_3$ nanoparticles is exposed to a pulse of UV light. d. Current modulation (ratio of the photocurrent over the dark current) as function of the Sn content of an ITO thin film exposed to a short illumination of UV light.

Under UV excitation, a large current modulation is observed, see Figure 3 and S5. A striking feature relative to the excitation of the band edge is the slow time response of the photocurrent, see Figure 3c. The decay time can be as long as 1h, which is the signature of the involvement of deep traps in the photoconductive process. Similar memory effect has already been observed for ZnO[49,50] nanoparticles and are now used as a strategy to activate the conduction of electron transport layer[51] in a solar cell device. The magnitude of the photoresponse tends to be reduced while the Sn content is increased. Again, the undoped nanoparticles present a significantly larger modulation than all doped nanoparticles, see Figure 3d, due to a much lower dark current. This observation combined with the weaker cross section of this material confirms that the free carrier density in the undoped material is low and certainly much weaker than the one determined using a Drude model. This suggests that the infrared peak observed for $In_2O_3$ nanoparticles has an intraband character, rather than plasmonic. Once the doping is introduced, this transition acquires a more collective character as it has been proposed in the case of self-doped HgS quantum dots.[52]

To support this claim, we graft polyoxometalates (POM) on the surface of the nanoparticles. It was recently demonstrated by Martinez *et al*[53] that POM can be used as agents to tune the doping magnitude of HgSe nanocrystals. The POMs used are a strongly oxidized (+VI) form of tungsten, and then behave as electron attractors. Once POMs are grafted on the surface of degenerately n-doped nanoparticles, they can strip electrons from the nanoparticles[53]. This strongly impacts the electron radial distribution[36] (Figure 4a) thanks to a post synthetic process. For doped forms of ITO, the grafting of the POMs has a severe effect of the linewidth of the plasmonic transition, see Figure 4b.

Initially ITO behaves as a Jellium[37] (Figure 4a), meaning that positive charges resulting from Sn dopants are fixed while free electrons are homogeneously spread all over the nanoparticle. After the POM grafting, the nanoparticle surface is electron deficient, which leaves positive impurities unscreened on the surface. This tends to enhance the electron impurity scattering and leads to the observed increase of the spectrum linewidth.

On the other hand, for $In_2O_3$, the infrared peak is neither shifted nor broaden, see c. In other words, the IR spectrum of the $In_2O_3$ nanoparticles with grafted POMs is just the sum of the absorption of the two materials. This reflects the fact that this transition is determined by the density of states of the semiconductor, rather than the electron gas.

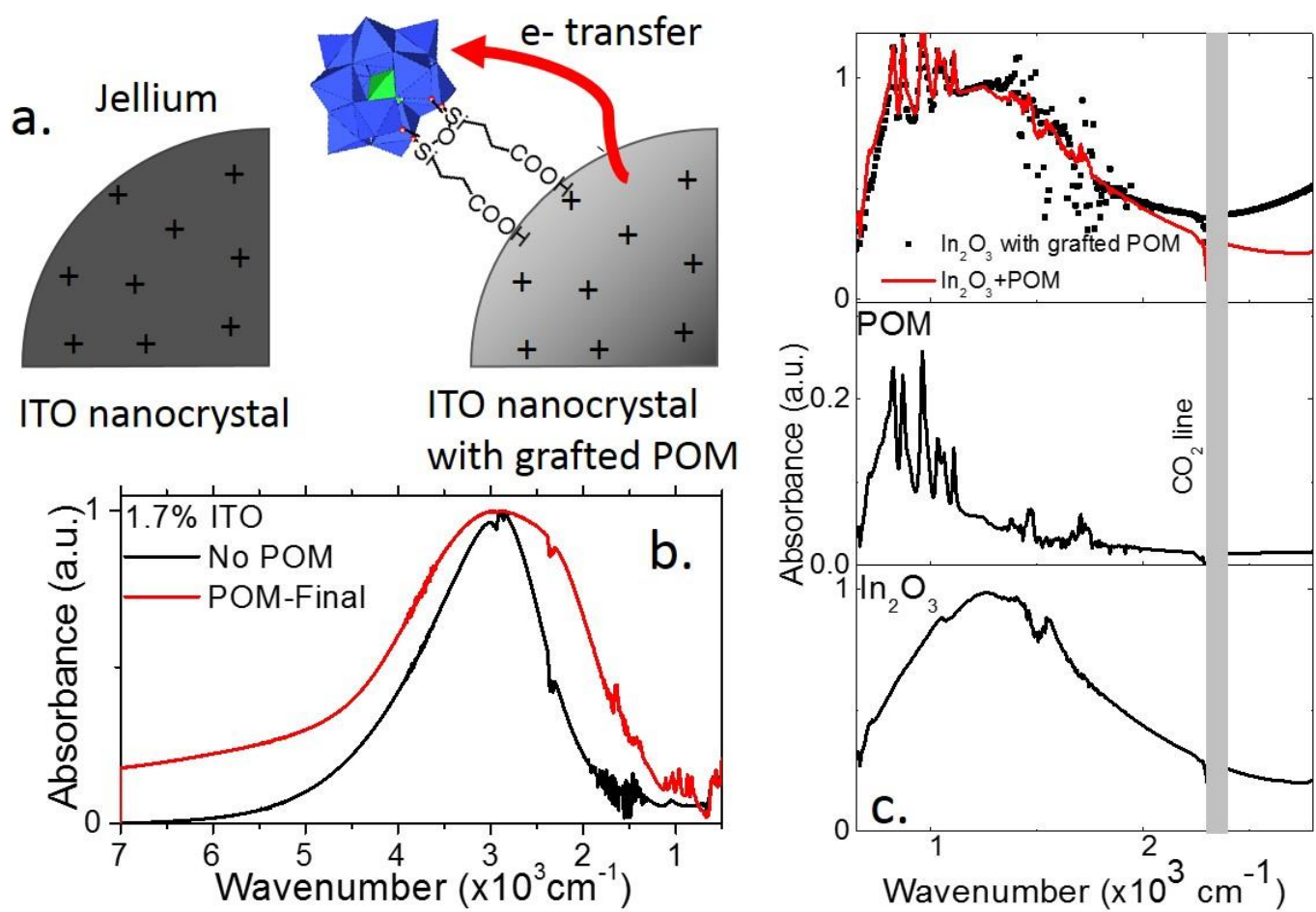

*Figure 4 a. Scheme of a non-functionalized nanoparticle where positive charges coming from Sn dopants are randomly dispersed in the nanocrystal, leading to a homogeneous distribution for the electron gas (in grey). In presence of electron attracting groups at the nanoparticle surface (here polyoxometalates), surface is electronically depleted and leave some of the positive charges unscreened. b. Infrared spectra of 1.7% doped ITO with and without POM grafted on the surface. c. Infrared spectrum of the $In_2O_3$ (bottom part), pristine POM (central part) and from the $In_2O_3$ nanoparticles grafted with POM (top part, scatter). The latter signal is well fitted (red curve) by an addition of the signal from the independent POM and $In_2O_3$ nanoparticles.*

There are several significant consequences to this observation beyond the obvious facts that undoped nanoparticles have lower conductivity and weaker absorption. First, reducing the doping and even the residual material non-stoichiometry will not allow to redshift the transition of ITO toward wavelengths longer than 9 μm. This may be a limitation for detection application in the low energy part of the 8-12 μm atmospheric transparency window. Secondly, if tuning of the intraband transition of $In_2O_3$ needs to be investigated, alternative paths such as change of confinement or material alloying need to be considered.

In the last part of the paper, we aim to discuss the potential of ITO nanocrystals as photoconductive material in the infrared range. Three ranges of detection have been considered : long-wave infrared for $In_2O_3$ nanoparticles, mid-wave infrared for 1.7% doped ITO and short-wave infrared for 10% doped ITO, see Figure 3a. Samples are cooled down to reduce their thermally activated carrier density, see Figure 5a, d and g. The responsivity achieved by the film of $In_2O_3$ nanocrystals reaches 40 µA/W in the long wavelength range, 4 µA/W in the mid infrared (1.7% doped ITO at 4.4 µm) and 14 µA/W at 1.55 µm for the 10 % Sn ITO.

The $In_2O_3$ nanoparticles present a dramatically slow photoresponse with turn-on and turn-off time close to 10 min, Figure 5b. All doped nanoparticles have a similar time responses, with a turn-on time above 10 s and a turn-off time of ≈1 min typically, see Figure 5e and h. Even if their dynamics are shorter than the photoconductive dynamics at the band edge, they remain extremely slow compared with other IR active colloidal materials in the same range of wavelengths. For example, with HgTe quantum dots, time response shorter than the µs are commonly reported.[54,55] This long time response actually suggests that the current modulation results from a bolometric effect. In other words, the change of carrier density induced by the light (*i.e.* the photocurrent) is negligible compared to the increase of thermally activated carrier density (*i.e.* the light-induced change of the dark current). We can conclude that the benefit of a stronger absorption brought by the plasmonic absorption is strongly balanced by the short lifetime of the photocarrier. Thus, to take full benefit of the larger plasmonic absorption, a fast extraction of the hot electrons will have to be implemented. Regarding the operating temperature of such films, we measured that the photocurrent remains a marginal modulation of the dark current, see Figure 5c, f and i. Only the undoped material achieve BLIP (background limited performances, defined here as the temperature when dark current becomes smaller that the photocurrent) operation at T<50K, see Figure 5c. This is the result of the large doping, required to achieve plasmonic absorption, but which comes at the price of a large dark current.

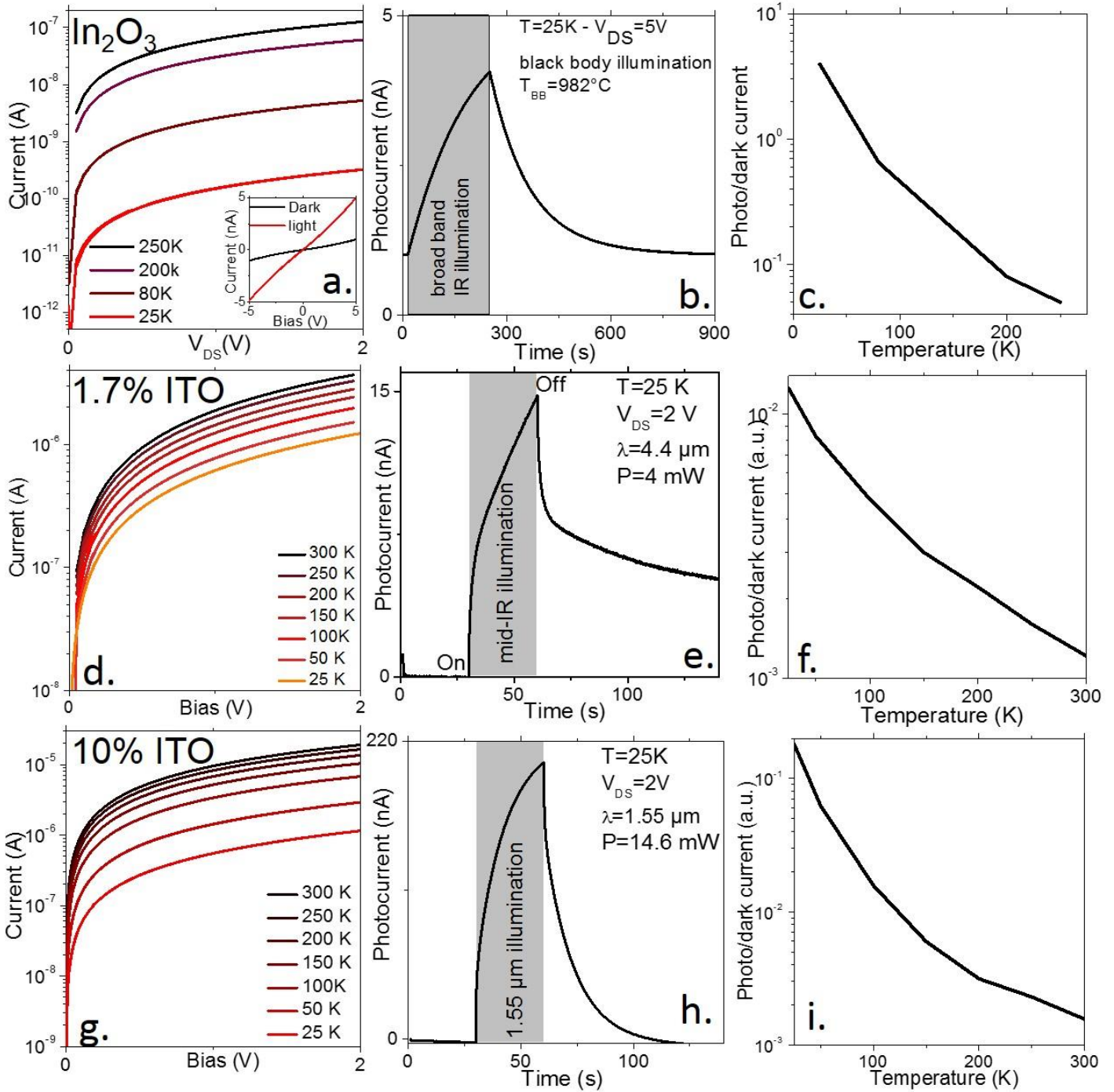

*Figure 5 I-V curves for a thin film of (a) $In_2O_3$ nanoparticles, (d) 1.7% doped ITO nanoparticles and (g) 10% doped ITO nanoparticles at various temperatures ranging from room temperature down to 25 K. The inset in (a) is the I-V curves at 25 K under dark condition and under illumination by a blackbody source ($T_{BB}$=982°C). Current as a function of time of a thin film of (b) $In_2O_3$ nanoparticles, (e) 1.7% doped ITO nanoparticles and (h) 10% doped ITO nanoparticles while exposed to (b) a pulse illumination by a blackbody source, (e) a short illumination of a quantum cascade laser operating at 4.4 µm and (h) a short illumination resulting from 1.55µm laser diode. Ratio of the photocurrent over the dark current as a function of the temperature for a thin film of (c) $In_2O_3$ nanoparticles exposed to the excitation from a blackbody source, (f) 1.7% doped ITO nanoparticles exposed to the excitation from a 4.4 µm QCL and (i) 10 % doped ITO nanoparticles exposed to the excitation from a 1.55 µm laser diode.*

## CONCLUSION

To summarize, we have tested the potential of ITO nanocrystals for infrared photoconduction. We have determined that the undoped $In_2O_3$ nanoparticles present an intraband absorption at 8-9 µm which transforms into a plasmonic transition as the Sn doping is introduced. The intraband character of this transition will make the tuning of the absorption toward longer wavelength more challenging. Doped nanoparticles of ITO present a cross section of a few

$10^{-13}$ cm$^2$ per nanoparticle, corresponding to an absorption coefficient above $10^5$ cm$^{-1}$. The material presents n-type conduction with a weak temperature dependence. Photoconduction in the infrared range can be obtained at low temperature and results from a bolometric effect. The responsivity achieved by thin films of ITO nanoparticle are in the few to few tens of µA/W$^{-1}$ range, which is not yet competitive with other materials such as PbS and HgTe. Future development will have to integrate a strategy to take advantage of the short living hot electrons.

## METHODS

### *List of chemicals*

Indium(III) acetate (Sigma-Aldrich, 99.99 %), tin(IV) acetate (Alfa Aesar), oleic acid (Sigma-Aldrich), olyel alcohol (Alfa Aesar, tech. 80-85%), ethanol absolute anhydrous (Carlo Erba, 99.9%), Chloroform (Carlo Erba), n-hexane (Carlo Erba), n-octane (SDS, 99%), acetic acid (Sigma-Aldrich, ⩾ 99%), N,N-dimethylformamide (DMF, Sigma Aldrich), triethyloxonium tetrafluoroborate (Et$_3$O$^+$ BF4$^-$, Sigma Aldrich), lithium perchlorate (LiClO$_4$, Sigma-Aldrich, 98%), polyethylene glycol (PEG, Mw=6 kg/mol) were used as received.

***Synthesis of indium oxide and Sn-doped indium oxide nanocrystals:*** In this study, the nanocrystals were synthesized with a two-step slow-injection method established by Jansons *et al.*[35] with minor modifications. In one three-neck flask, a desired composition of indium acetate and tin (IV) acetate (5 mmol in total) was mixed with oleic acid (10 mL). The mixture was heated at 150 °C under Ar until all the powders are fully dissolved (typically 1 hour). The obtained metal oleate solution (0.5M) was light yellow in color. In another flask, 13 mL of oleyl alcohol was heated to 305 °C under Ar. Then 1 mL of the as-prepared metal oleate solution was injected to the oleyl alcohol bath using a syringe-pump at a rate of 0.2 mg/mL. After the injection, the reaction was baked at 305 °C under Ar for 20 min before cooled down with air flux. The obtained nanocrystals were precipitated with ethanol and redispersed in chloroform for 3 times and finally dispersed in chloroform for storage. The nominal doping of Sn in nanocrystal was determined by the Sn/(In+Sn) ratio in the mixed precursor. In this manner, we synthesized a series of nanocrystals with 0% (In$_2$O$_3$), 1.7%, 3%, 5% and 10% of Sn doping.

**POM-COOH (TBA)$_3$[PW$_{11}$O$_{39}${O(SiC$_2$H$_4$COOH)$_2$}] synthesis** [56]: K$_7$[PW$_{11}$O$_{39}$] (0.64 g, 0.2 mmol) was dissolved in a water/acetonitrile mixture (30 mL, 1:2). A 1 M HCl aqueous solution was added drop by drop until an apparent pH equals to 3. The solution was cooled in an ice bath and the Si(OH)$_3$(CH$_2$)$_2$COONa (0.476 mL, 0.8 mmol) was inserted. The 1 M HCl solution was added drop by drop again to reach pH$_{app}$=2. After an overnight reaction, TBABr (0.26 g, 0.8 mmol) was added and the solution concentrated with a rotary evaporator to make precipitate the product. The oily compound obtained was dissolved in the minimum of acetonitrile then precipitated again with an excess of ether. A sticky solid was recovered by centrifugation and washed thoroughly with ether to obtain a white powder (0.6 g, 82%). The dried compound was found to be partially deprotonated (cf. IR and EA analysis) with the exact general formula (TBA)$_{3,4}$[PW$_{11}$O$_{39}${O(SiC$_2$H$_4$COOH$_{0,8}$)$_2$}].

$^1$H NMR (400 MHz, CD$_3$CN) : δ(ppm) 3.14 (m, 24H), 2.53 (m, 4H), 1.65 (m, 24H), 1.41 (sex, $^3$J(H,H)=7.5 Hz, 24H), 1.05 (m, 4H), 1.01 (t, $^3$J(H,H)=7.5 Hz, 36H), 0.90 (m, 4H) ;

$^{31}$P NMR (121 MHz, CD$_3$CN) δ (ppm) -12.28 ;

IR (KBr pellet) : δ =2963 (s), 2935 (m), 2874 (w), 1710 (s), 1623 (w), 1483 (s), 1471 (s), 1420 (w), 1381 (m), 1112 (vs), 1064 (vs), 1052 (s), 1036 (s), 964 (vs), 870 (vs), 824 (vs)

MS (ESI-), m/z (%) : calcd for W$_{11}$PSi$_2$O$_{44}$C$_6$H$_{10}$ : 965.41 [M]$^{3-}$ ; found : 965.42 (100) ; calcd for W$_{11}$PSi$_2$O$_{44}$C$_{22}$H$_{46}$N : 1569.26 [M+TBA]$^{2-}$ ; found : 1569.27 (50)

Elemental Analysis calcd (%) for $C_{60.4}H_{132}N_{3.4}O_{44}Si_2PW_{11}$ : C 19.50, H 3.55, N 1.28; found : C 19.51, H 3.50, N 1.24

**POM grafting on ITO nanoparticles**: We used a two-step ligand exchange method with the removal of organic ligands followed by the functionalization with POM. Firstly 1 mL of NCs in hexane (10 mg/mL) was mixed with 1 mL of triethyloxonium tetrafluoroborate in DMF (20 mg/mL) vigorously with vortex and sonication. After 5 min, the NCs were transferred from nonpolar phase (hexane) to the polar one (DMF) on the bottom, due to the stripping of their organic ligands. Then, the bare NCs in DMF were precipitated with toluene and redissolved in 1 mL of POM in DMF (20 mg/mL) for functionalization. After complete mixing, the POM-capped NCs were precipitated with toluene to remove the POM excess and were redispersed in DMF for tests.

**Ligand exchange for transport measurements:** To improve electron transport, the long organic ligands of nanocrystals were replaced by short ones using an on-film ligand exchange method. In more details, we first drop-casted the nanocrystal solution onto a substrate. Once the substrate was dried, it was dipped into a solution of acetic acid in ethanol (0.5 wt%) for 60 s then in pure ethanol for 30 s. The substrate was then annealed at 250 °C for 15 min to finish one round of ligand exchange. The procedure was repeated once before transport measurements.

**Material characterization:** Absorption spectra are acquired using a Jasco V730 spectrometer for the UV visible part, while a Thermo Fischer IS50 Fourrier transform infrared spectrometer in ATR configuration is used in the IR range. For the determination of the absorption cross section, a film of nanoparticles has been deposited onto a double side polished Si wafer. Its thickness is determined using a Dektak 150 profilometer. The absorption coefficient is given by $\frac{A \cdot ln10}{t \cdot f}$ with A the absorption, t the film thickness, and f the film volume fraction taken equal to 0.64 which correspond to a randomly close pack film. The cross section per particle is simply obtained by multiplying the absorption coefficient by the nanoparticle volume, assuming a spherical shape. For Transmission electron microscopy, we used a JEOL 2010. The grids were prepared by a drop cast of dilute solution of nanocrystals dispersed in hexane and degassed overnight under secondary vacuum. X-ray diffraction pattern is obtained by drop casting a solution of nanocrystals on a Si wafer. The diffractometer is a Philips X'Pert, based on the emission of the Cu Kα line operated at 40 kV and 40 mA current.

**Si/SiO$_2$ substrate for electrodes:** The surface of a Si/SiO$_2$ wafer (400 nm oxide layer) is cleaned by sonication in acetone. The wafer is rinsed with isopropanol and finally cleaned using an O$_2$ plasma. AZ 5214E resist is spin-coated and baked at 110°C for 90 s. The substrate is exposed under UV through a pattern mask for 2 s. The film is further baked at 125°C for 2 min to invert the resist. Then a 40 s flood exposure is performed. The resist is developed using a bath of AZ 326 for 32 s, before being rinsed in pure water. We then deposit a 5 nm chromium layer and 80 nm gold layer using a thermal evaporator. The lift-off is performed by dipping the film in acetone for 1 h. The electrodes are finally rinsed using isopropanol and dried by an air flow. The electrodes are 2.5 mm long and spaced by 20 µm. These electrodes are used for DC measurements (IV curves and transistor measurements).

**Electrolyte gating**: For electrolyte gating, we first mix in a glove box 0.5 g of LiClO$_4$ with 2.3 g of PEG ($M_W$ = 6 kg.mol$^{-1}$). The vial is heated at 170°C on a hot plate for 2 h until the solution gets clear. To use the electrolyte, the solution is warmed around 100°C and brushed on the top of the ITO nanoparticle film.

**Electrical measurements**
**DC transport**: The sample is connected to a Keithley 2634b which applied bias and measured current. For measurement under illumination three sources has been used: a UV flash light at 365 nm, a 1.55µm laser diode and a quantum cascade laser operating at 4.4 µm. For measurement as a function of temperature, the sample is mounted on the cold finger of a cryostat and the sample is biased using a Keithley 2634b. The current is acquired while the temperature of the sample is cooled down.

**Transistor measurements**: The sample is connected to a Keithley 2634b which sets the drain bias ($V_{DS}$ = 20 mV for doped ITO, $V_{DS}$ = 200 mV for $In_2O_3$), controls the gate bias ($V_{GS}$) between -2 and +2 V with a step of 1 mV and measures the associated currents $I_{DS}$ and $I_{GS}$. All measurement are conducted in room condition (temperature and pressure).

Supporting Informations

The Supporting Information is available free of charge on the ACS Publications website at DOI:
Size determination of nanoparticles, X-ray diffractogram, Tauc plot, process and effect of POM grafting, field effect transistor measurements and photoconduction under UV illumination

ACKNOWLEDEGMENTS

We thank C. Delerue for valuable discussion on the effect of the charge distribution. EL thanks the support ERC starting grant blackQD (grant n° 756225). We acknowledge the use of clean-room facilities from the "Centrale de Proximité Paris-Centre". This work has been supported by the Region Ile-de-France in the framework of DIM Nano-K (grant dopQD). This work was supported by French state funds managed by the ANR within the Investissements d'Avenir programme under reference ANR-11-IDEX-0004-02, and more specifically within the framework of the Cluster of Excellence MATISSE and also by the grant Nanodose and IPER-Nano2. JQ thanks the Chinese Scholar council for PhD grant while NG and JR thank Nexdot for post doctorate funding.